\documentclass[letter,twocolumn]{jpsj5}

\title{Anomalous Temperature Dependence of the Superfluid Density Caused by Dirty-to-Clean Crossover in FeSe$_{0.4}$Te$_{0.6}$ Single Crystals}

\author{Hideyuki Takahashi$^{1,3}$, Yoshinori Imai$^{1,3}$, Seiki Komiya$^{2,3}$, Ichiro Tsukada$^{2,3}$, Atsutaka Maeda$^{1,3}$}

\inst{$^{1}$Department of Basic Science, the University of Tokyo, 3-8-1 Komaba, Meguro-ku, Tokyo 153-8902, Japan \\
$^{2}$Central Research Institute of Electric Power Industry, 2-6-1 Nagasaka, Yokosuka, Kanagawa 240-0196, Japan\\
$^{3}$Transformative Research Project on Iron Pnictides (TRIP), JST, Japan}

\abst{
We report microwave surface impedances of FeSe$_{0.4}$Te$_{0.6}$ single crystals measured at 12, 19, and 44 GHz. The penetration depth exhibits a power law behavior, $\delta \lambda_L=\lambda_L (T)-\lambda_L (0) \propto CT^n$ with an exponent $n\simeq 2$, which is considered to result from impurity scattering. This behavior is consistent with $s\pm$-wave pairing symmetry. The temperature dependence of the superfluid density largely deviates from the behavior expected in the BCS theory. We believe that this deviation is caused by the crossover from the dirty regime near $T_c$ to the clean regime at low temperatures, which is supported by the rapid increase of the quasiparticle scattering time obtained from the microwave conductivity. We also believe that the previously published data of the superfluid density can be interpreted in this scenario.
}

\begin{document}
\maketitle

Iron-based superconductors have attracted much attention since the discovery of LaFeAsO$_{1-x}$F$_{x} $ with $T_c=28~\mathrm{K}$~\cite{Kamihara2008}.
These compounds are expected to have an unconventional pairing mechanism because they contain Fe, one of the most familiar ferromagnetic elements. Indeed, there have been many studies on various pairing symmetries of iron-based superconductors~\cite{NatPhys6.645, PhysRevLett.101.057003, PhysRevLett.101.087004, PhysRevLett.104.157001}. 
The penetration depth studies have also suggested that the details of the pairing symmetry depends on the materials~\cite{PhysRevLett.102.017002, PhysRevLett.102.207001, PhysRevLett.102.147001, PhysRevLett.104.087005}.\\
\indent However, the estimates of the value of the superconducting gaps are inconsistent among different groups even in the same compounds~\cite{NatPhys6.645}. One reason is that the temperature dependence of the superfluid densities $n_s(T)$ in some compounds cannot be understood in the framework of the BCS model. Typical example is $n_s(T)$ in FeSe$_{0.4}$Te$_{0.6}$, which we focus in this paper. FeSe$_{0.4}$Te$_{0.6}$~\cite{PhysRevB.78.224503} ($T_c=14$~K) (the so-called 11 system) has the simplest crystal structure (PbO-type) among all iron based superconductors. This compound does not contain arsenic, which is common in most iron-based superconductors, but the electronic structure of this system is similar to that of the FeAs layers~\cite{PhysRevB.78.134514}.
It was suggested that $n_s(T)$ in Fe$_{1.03}$Se$_{0.37}$Te$_{0.63}$ can be fitted by a two gap model that takes into account interband scattering~\cite{PhysRevB.81.180503}. However, the model is based on the assumption that the the sample is within the clean limit.
In the 11 system, unlike ordinary metals, the resistivity of the material shows a small temperature dependence and has a relatively large value in the normal state, $\rho(15~\mathrm{K})\simeq 0.5~\mathrm{m\Omega cm}$~\cite{PhysRevB.79.094521,PhysRevB.80.092502,JPSJ.79.084711}. Thus, it is possible that there are unusual characteristics of quasiparticle scattering that affect the superconductivity and that produce some anomalous superconducting features. We believe that the microwave conductivity measurement is needed to evaluate the quasiparticle scattering in the superconducting state. In this paper, we present the results of surface impedance measurements in FeSe$_{0.4}$Te$_{0.6}$ single crystals, from which we extracted the microwave conductivity $\sigma$ and $n_s$. Then, we show that the crossover from the dirty regime near $T_c$ to the clean regime at low temperatures can affect $n_s(T)$. We believe that the scenario is applicable to many other iron-based superconductors which have the "dirty" nature in the normal state.\\
\indent
The single crystals of FeSe$_{0.4}$Te$_{0.6}$ were grown by the Bridgman method~\cite{PhysRevB.79.094521}. The starting materials were grains of Fe (purity 3N), Se (purity 5N), and Te (purity 5N).
These materials prescribed in the molar ratio of Fe:Se:Te=1:0.6:0.4 were sealed in an evacuated quartz tube.
Consequently, the doubled-wall sealed quartz tube was heated at 650$^\circ\mathrm{C}$ for 100 h and successively at 1000$^\circ\mathrm{C}$ for 36 h and then cooled down to 650$^\circ\mathrm{C}$ for more than 200 h and then eventually down to room temperature by furnace cooling.
As described in previous papers~\cite{PhysRevB.80.092502,JPSJ.79.084711}, as-grown crystals were annealed at 450$^\circ\mathrm{C}$ for 200h in vacuum to improve the crystal quality.
Although there are excess iron atoms that partially occupy the interstitial sites of the (Te, Se) layers in the 11 compounds~\cite{PhysRevB.79.094521}, the metallic behavior in $\rho(T)$ shows a low amount of excess irons in our samples.
The transition temperature determined from dc magnetic susceptibility is $T_c\sim 14~\mathrm{K}$ with the width $\Delta T<0.5~\mathrm{K}$, which is rather sharp.\\
\indent
The surface impedance $Z_s=R_s-iX_s$, where $R_s$ is the surface resistance and $X_s$ is the surface reactance, was measured by a cavity perturbation technique.
We used three kinds of cylindrical oxygen-free Cu cavity resonators, operated in the TE$_{011}$ mode at 12, 19 and 44~GHz, which have a quality factor, $Q\sim60000$ (12~GHz, 19~GHz), 26000 (44~GHz), respectively. 
A piece of crystals is mounted on a sapphire rod and is placed in the antinode of the microwave magnetic field $H_{\omega}$. $H_{\omega}$ is parallel to the $c$-axis, so that the shielding current flows in the $ab$ planes. In this technique, one measures the changes of $Q$ and the resonant frequency $f$ of the cavity, which are caused by introducing a sufficiently small sample. The shifts in the inverse of $Q$ and the resonant frequency, expressed by $\Delta(\frac{1}{Q})=\frac{1}{Q_s}-\frac{1}{Q_b}$, and $\Delta f=f_s-f_b$ (where the indices $s$ and $b$ indicate the values with and without the sample, respectively), are proportional to $R_s$ and $X_s$, respectively. The absolute values of $Z_s$ were obtained by assuming the Hagen-Rubens limit ($\omega\tau\ll 1$), where $R_s=X_s=(\mu_0 \omega \rho/2)^{\frac{1}{2}}$ above $T_c$ ($\omega=2\pi f$ is the angular frequency, $\tau$ is the quasiparticle scattering time, and $\mu_0$ is the permeability in vacuum).
 
For the case of local electrodynamics, we can extract $\sigma$ from $Z_s$ using the relation $Z_s=(-i\mu_0\omega/\sigma)^{\frac{1}{2}}$. 
In the two-fluid model, $\sigma$ is expressed as:
\begin{equation}
\sigma=\sigma_1+i\sigma_2=\frac{n_n e^2\tau}{m^*}\frac{1}{1-i\omega\tau}+i\frac{n_s e^2}{m^* \omega},
\label{Drudesigma}
\end{equation}
where we assume the Drude-like normal fluid, $e$ is the electric charge, and $n_n/m^*$ and $n_s/m^*$ are the normal fluid and the superfluid density over the effective mass, respectively. 
At low temperatures and at low frequencies,
 the assumption $\sigma_1\ll \sigma_2 $ is valid, and one obtains the relation:
\begin{equation}
X_s=\mu_0 \omega \lambda_L ,
\label{Xs_lambda}
\end{equation}
where $\lambda_L$ is the London penetration depth. 
Since $\lambda_L$ is related to $n_s$ via the London equation $\lambda_L^{-2}=\mu_0 n_s e^2/m^*$, $\lambda_L(T)$ at low temperatures will give us information about the superconducting gap structure, particularly the presence or the absence of nodes in the gap.

\begin{figure}[tb]
\includegraphics[width=7.5cm]{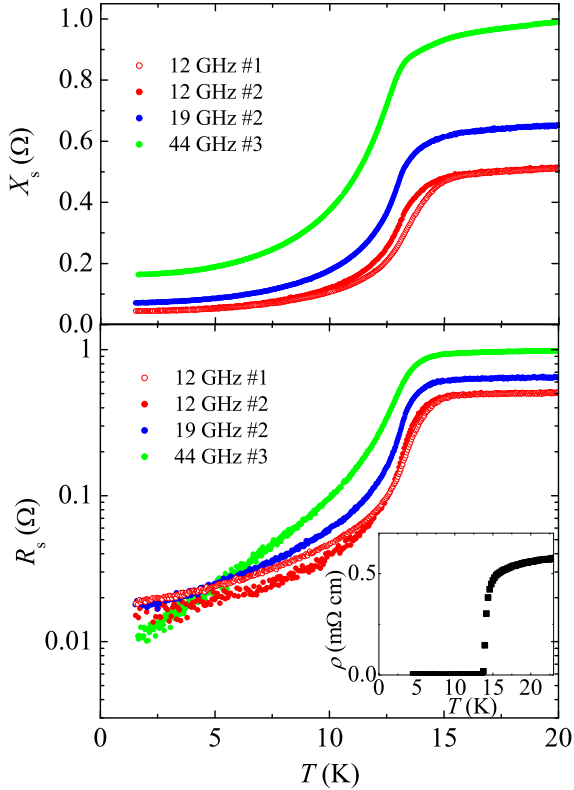}
\caption{(color online) The temperature dependence of the microwave surface impedance of FeSe$_{0.4}$Te$_{0.6}$ single crystals, $Z_s=R_s-iX_s$, at 12, 19, and 44~GHz, respectively. 
Crystal \#1 (with dimensions $1.0\times1.0\times0.1~\mathrm{mm}^3$) shows sharper superconducting transition than crystal \#2 ($0.7\times0.5\times0.1~\mathrm{mm}^3$ for 12 and 19~GHz) and crystal \#3 ($0.5\times0.3\times0.05~\mathrm{mm}^3$ for 44~GHz). The inset shows $\rho(T)$ near $T_c$.}
\label{Zs_all}
\vspace{-5mm}
\end{figure}
Figure \ref{Zs_all} shows the temperature dependence of $Z_s$ at the three frequencies.
In general, $R_s$ increases as the frequency increases.
However, in our samples, crystal \#3 at 44~GHz measurement shows the lowest $R_s$ at 1.6~K.
This result indicates that the main contribution to the residual surface resistance $R_s^{res}$ has an extrinsic origin.
The surface impedance measurements are susceptible to defects on the sample surface.
For layered conductors in general, the presence of delaminated edges on the sample is reported to cause excess loss of microwave power~\cite{PhysRevB.82.094520}.
Thus, we expect that the thicker samples to show larger values of $R_s^{res}$.
In fact, crystal \#3, which was cleaved from crystal \#2 after the 12~GHz and 19~GHz measurements, is the thinnest among the samples and has the lowest $R_s^{res}$, which agrees with the above expectation. 

\begin{figure}[tb]
\begin{center}
\includegraphics[width=7.5cm]{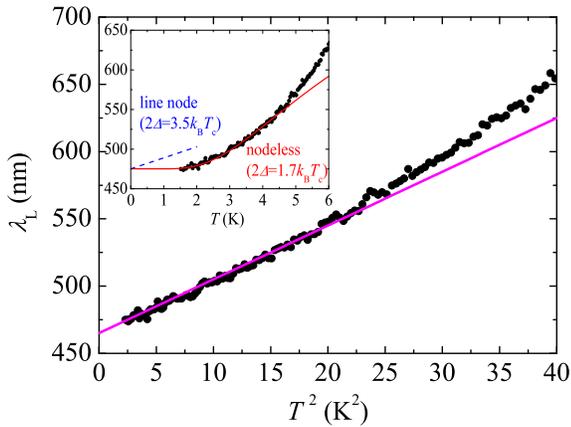}
\caption{(color online) The temperature dependence of the magnetic penetration depth $\lambda_L$. The purple solid line represents the $T^2$ behavior. The inset shows the fit to nodeless $s$-wave (red solid line) and the behavior of $d$-wave with lines of nodes (blue dashed line). }
\label{lam}
\end{center}
\vspace{-6mm}
\end{figure}

Figure \ref{lam} shows $\lambda_L (T)$ for crystal \#2 at low temperatures.
The absolute value of $\lambda_L$ at 0 K is $\lambda_L(0)\simeq470~\mathrm{nm}$, which is in fair agreement with the other measurements~\cite{PhysRevB.81.180503,PhysRevB.82.184506}.
The penetration depth is found to behave in a power-law manner, $\delta \lambda_L\equiv\lambda_L (T)-\lambda_L (0) \propto CT^n$, with the exponent $n\simeq 2$.
This behavior is also consistent with the previous electrodynamic study in MHz region~\cite{PhysRevB.81.180503,PhysRevB.82.184506}. 
In general, a power-law temperature dependence implies the presence of low-energy quasiparticle excitations. 
In this case the quasiparticle density of states (DOS) behaves as $D(E) \propto E^n$ close to the Fermi level, where $E$ is the quasiparticle energy.
At first sight, our result might seems to be inconsistent with a very flat DOS observed in an STM study~\cite{Science.328.474}.
However, it has been shown that in a two-band superconductor with $s\pm$-wave symmetry with nonmagnetic impurities, the behavior of $\lambda_L$ at low temperature is essentially the same as in a conventional $s$-wave superconductors with a considerable amount of magnetic impurities~\cite{PhysRevB.79.140507}. 
Thus, $s\pm$-wave symmetry is plausible because the crystals contain low excess iron as the magnetic impurities.

\begin{figure}[tb]
\begin{center}
\includegraphics[width=7.5cm]{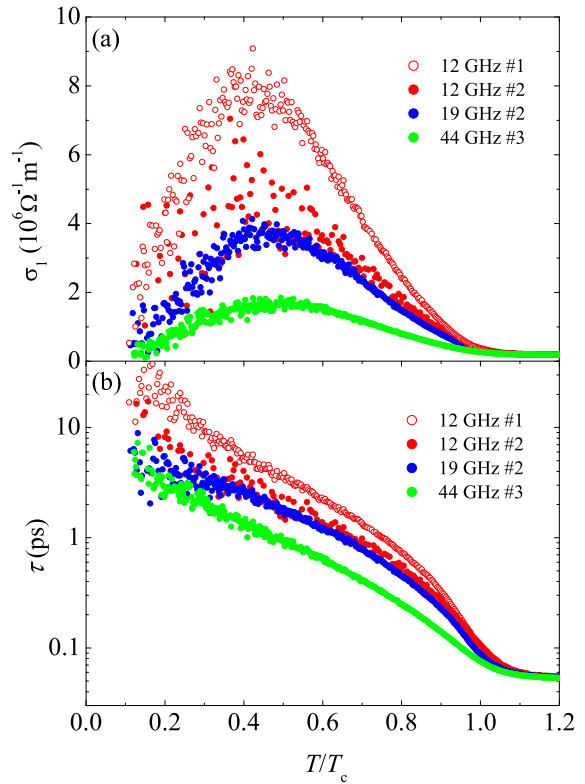}
\end{center}
\vspace{-5mm}
\caption{(color online) (a) The temperature dependence of the real part of the microwave conductivity. (b) The temperature dependence of the quasiparticle scattering time in the superconducting state.}
\label{sigtau}
\vspace{-4mm}
\end{figure}

Figure \ref{sigtau}(a) shows $\sigma_1 (T)$ in the superconducting state. 
By subtracting $R_s^{res}$ from the raw data of $R_s$ before calculating $\sigma$, we can avoid the influence of surface defects and precisely discuss the quasiparticle dynamics.
It should be noted that a small error in estimating $R_s^{res}$ does not change the essential feature of $\sigma_1(T)$ at large. 
In all samples, a considerable enhancement of $\sigma_1$ was observed below $T_c$. Although the magnitude of the peak is somewhat different between the two samples (see the 12~GHz data), it tends to decrease with increasing frequency.
It is well known that in the clean limit of the BCS superconductors, $\sigma_1(T)$ has a coherence peak.
However, the peaks observed in our $\sigma_1(T)$ data are much larger and broader than the coherence peak.
Such peaks were also observed in the cuprate superconductors~\cite{PhysRevLett.68.2390,PhyicaC203.315,PhysRevB.60.1349} and in other iron-based superconductors~\cite{PhysRevLett.102.017002,PhysRevLett.102.207001}.
In those systems, inelastic scattering is dominant in the normal state.
Below $T_c$, inelastic scattering is suppressed because the quasiparticle DOS near the Fermi level decreases by the emergence of the superconducting gap, giving rise to the peak in $\sigma_1(T)$ below $T_c$. 
Conversely, an enhancement of $\sigma_1$ indicates that inelastic scattering is dominant above $T_c$.

Since the coherence effect is not important in this system, we used the two-fluid model [see Eq.~(\ref{Drudesigma})] to extract $\tau$ and $n_s$ as follows:
\begin{equation}
\omega\tau=\frac{\tilde{\sigma_1}}{1-\tilde{\sigma_2}},
\end{equation}
\begin{equation}
\frac{n_s(T)}{n_s(0)}=\tilde{\sigma}_2-\frac{\tilde{\sigma}_1^2}{1-\tilde{\sigma}_2},
\label{fs}
\end{equation}
where we assume that all carriers condense at $T=0~\mathrm{K}$. We also introduced the dimensionless conductivity $\tilde{\sigma}=\tilde{\sigma_1}+i\tilde{\sigma_2}=\mu_0\omega\lambda_L^2(0)(\sigma_1+i\sigma_2)$. 
Figure \ref{sigtau}(b) shows the temperature dependences of $\tau$.
As anticipated, we found the rapid increase in $\tau$ in all samples below $T_c$.
The magnitude of the peak in $\sigma_1(T)$ reflects the degree of impurity scattering.
In crystal \#1, which has a larger peak than in crystal \#2, $\tau$ increases more rapidly than in \#2.
The quasiparticle scattering time $\tau$ in crystal \#1 reaches more than 10~ps far below $T_c$.
This value is two orders of magnitude larger than that in the normal state and is comparable to that in LiFeAs in the superconducting state~\cite{JPSJ.80.013704}.

We find that $\tau$ depends on frequency.
As frequency decreases, $\tau$ increases more rapidly in the superconducting state.
Although we do not completely understand the origin of this frequency dependence, it might be related to the energy dependence of the DOS near the Fermi level. It is consistent with the observation that the frequency dependence is more profound at higher temperatures, where the number of thermally excited quasiparticles increases.

\begin{figure}[tb]
\begin{center}
\includegraphics[width=7.5cm]{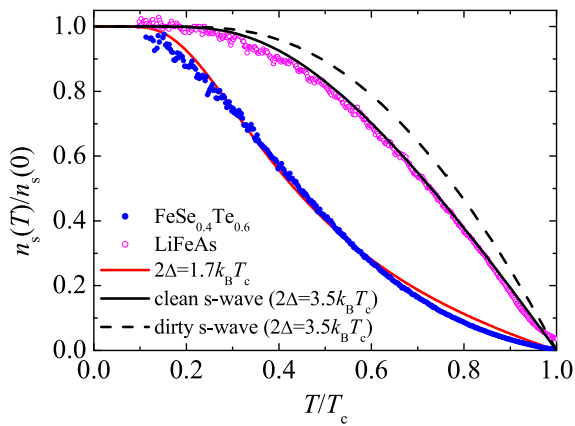}
\vspace{-3mm}
\caption{(color online) The temperature dependences of the superfluid density $n_s$ in FeSe$_{0.4}$Te$_{0.6}$ and LiFeAs~\cite{JPSJ.80.013704}. For clarity, only the 19~GHz data for crystal \#2 are shown. The black solid (dashed) line is a curve expected for a conventional superconductor with $2\Delta=3.5~k_B T_c$ in the clean (dirty) limit. The red solid curve is a fit to clean $s$-wave from 0.3~$T_c$ to $T_c$.}
\label{ns}
\end{center}
\vspace{-7mm}
\end{figure}

Next, we discuss $n_s (T)$.
When Eq.~(\ref{Xs_lambda}) is valid, $n_s(T)$ is given by $n_s(T)/n_s(0)=\lambda_L(0)^2/\lambda_L(T)^2=X_s(0)^2/X_s(T)^2$. However, there was some deviation from Eq.~(\ref{Xs_lambda}) at the highest measurement frequency, 44~GHz, and in the temperature range of 6-14~K, since the normal fluid contribution to $X_s$ is not negligible. In this case, we have to consider the quasiparticle dynamics by analyzing $\sigma$ data using Eq.~(\ref{fs}).
Figure \ref{ns} shows $n_s(T)/n_s(0)$ in crystal \#2, which is very much different from that of the BCS superconductors.
It is clear that this is convex downward in a wide temperature range. 
Near $T_c$, $n_s$ increases very slowly with decreasing temperature, 
but this is irrelevant to the sample inhomogeneity such as the spatial distribution of Se content, because our samples show a rather sharp superconducting transition.
For the conventional $s$-wave superconductors, such temperature dependence does not appear whether they are in the clean limit or the dirty limit~\cite{Tinkham}.
This behavior is sometimes interpreted as a characteristic of multigap superconductors having at least one very small gap.
If we try to fit the data by a two-gap model, where $n_s(T)=xn_{s1}(T)+(1-x)n_{s2}(T)$ ($0<x<1$), we get the gap value $2\Delta\simeq1.7~k_B T_c$ for both gaps, which is much smaller than both the values $2\Delta=3.5~k_B T_c$ expected in the BCS theory and the value $2\Delta=3.6~k_B T_c$ measured by STM~\cite{Science.328.474}.

To explain this peculiar temperature dependence, we have to clarify whether FeSe$_{0.4}$Te$_{0.6}$ is a clean superconductor or not by comparing the electric mean free path $l$ with the coherence length $\xi$. We can calculate $l=v_F \tau$ with the Fermi velocity $v_F$ measured by the angle-resolved photoemission spectroscopy (ARPES) on FeSe$_{0.5}$Te$_{0.5}$~\cite{PhysRevLett.104.097002}. In ARPES, a very small $v_F$ is observed for the outer hole pocket ($\alpha_3$) centered at the É° point and the electron pocket at the M point. With $v_F\simeq 1.4\times10^{-4}~\mathrm{m/s}$ for the $\alpha_3$ pocket, $l$ is about 7~{\AA} just above $T_c$. This is half of the coherence length, $\xi\sim15$~{\AA}, estimated from an upper critical field measurement~\cite{PhysRevB.82.184506,JJAP.49.023101}. Thus, the sample is in the dirty regime near $T_c$. As the temperature decreases, $l$ extends with the rapid increase in $\tau$, and $l>100$~{\AA} at 0.5~$T_c$, which shows that the sample is in the clean limit.

Therefore, the behavior of $n_s(T)/n_s(0)$ can be understood as follows: only a small amount of quasiparticle collapses into the condensate near $T_c$ because $l\simeq\xi$. 
At sufficiently low temperatures, the sample enters the clean limit, where $n_s$ increases with a massive loss of conductivity spectral weight, resulting in an $n_s(T)/n_s(0)$ with a positive curvature.
To summarize, it is important for $n_s(T)/n_s(0)$ in FeSe$_{0.4}$Te$_{0.6}$ that there is a crossover from the dirty regime to the clean regime in the superconducting state, with decreasing temperature.
This interpretation can be also applied to other iron-based superconductors such as the doped compounds Ba(Fe,Co)$_2$As$_2$~\cite{PhysRevLett.102.127004} and (Ba,K)Fe$_2$As$_2$~\cite{PhysRevLett.102.207001}, where similar temperature dependence is often observed.
Furthermore, the report on (Ba,K)Fe$_2$As$_2$ which showed that $n_s(T)$ changes dramatically with a small impurity concentration~\cite{PhysRevLett.102.207001} can be understood as a result of the increase of $\xi/l$ near $T_c$. 
On the other hand, as for the stoichiometric LiFeAs which shows $\xi/l<1$ even above $T_c$, $n_s(T)$ can be fitted by a simple two-gap model~\cite{JPSJ.80.013704} because the samples are in the clean regime at all temperatures below $T_c$. 

In conclusion, we have measured $Z_s$ of FeSe$_{0.4}$Te$_{0.6}$ single crystals.
The quadratic temperature dependence of $\lambda_L$ indicates the presence of impurity scattering.
We believe that $s\pm$-wave pairing is the most plausible possibility to consistently explain this behavior.
In the superconducting state, an enhancement of $\sigma_1$ caused by the rapid increase of $\tau$ was observed.
The temperature dependence of superfluid density $n_s(T)/n_s(0)$ shows a positive curvature in most of the temperature regime measured because the crystals are not in the clean limit near $T_c$ and only become clean below $T_c$.
This behavior indicates that one should carefully assess the relation between $l$ and $\xi$ in discussing $n_s(T)$ in iron-based superconductors.

\appendix

We appreciate Noriyuki Nakai and Tetsuo Hanaguri for the fruitful discussions.



\hyphenation{Post-Script Sprin-ger}


\end{document}